\documentclass[%
 preprint,
superscriptaddress,
 amsmath,amssymb,
 aps,
 onecolumn,
 notitlepage
]{revtex4-2}

\usepackage{epstopdf, epsfig}
\usepackage{natbib}
\usepackage{amsmath}

\usepackage{graphicx}
\usepackage{dcolumn}
\usepackage{bm}
\usepackage[utf8]{inputenc}
\usepackage{subfig}
\usepackage[normalem]{ulem}

\usepackage{tikz}
\usepackage{placeins}
\usepackage[european resistor, european voltage, european current]{circuitikz}
\usetikzlibrary{arrows,shapes,positioning}
\usetikzlibrary{decorations.markings,decorations.pathmorphing,
decorations.pathreplacing}
\usetikzlibrary{calc,patterns,shapes.geometric}

\newcommand\Sh{\mathcal Sh}

\usepackage{color}

\begin{document}

\title{Numerical simulation of non-Brownian suspensions: Normal viscosity and Viscous resuspension}


\author{William Ch\`{e}vremont}
\affiliation{%
Univ. Grenoble Alpes, CNRS, Grenoble INP, LRP, 38000 Grenoble, France
}%


\author{Bruno Chareyre}
\email{bruno.chareyre@grenoble-inp.fr}
\affiliation{
 Univ. Grenoble Alpes, CNRS, Grenoble INP, 3SR, 38000 Grenoble, France
}%
\author{Hugues Bodiguel}
\affiliation{%
Univ. Grenoble Alpes, CNRS, Grenoble INP, LRP, 38000 Grenoble, France
}%

\date{\today}

\begin{abstract}
Normal stresses in sheared suspensions of non-Brownian particles are obtained from numerical simulations in the viscous regime. The stresses are determined in homogeneous shear of non-buoyant particles and by analyzing shear-induced resuspension of buoyant particles in the framework of the suspension balance model (SBM). The consistency of both approaches indicates that the SBM describes the steady state properly. The results are in agreement with some earlier empirical expressions of the normal viscosity coefficients in limited ranges of particle volume fractions, but they appear to overestimate the stresses in the semi-dilute regime (solid fraction in the range 25-35\%). New expressions are proposed for. We show that these discrepancies can be due to inertial effects which appear at rather low particle Reynolds number. The results also highlight that the normal stress anisotropy depends on the volume fraction.
\end{abstract}

\pacs{Valid PACS appear here}


\maketitle

\section{\label{sec:introduction}Introduction}

Suspensions of solid particles dispersed in a liquid are ubiquitous in nature and in technology. Describing their flow properties is therefore important issue, but despite intense research efforts, the global understanding is not yet achieved, due to the complexity of the problem. For high enough Peclet numbers, Brownian motion has negligible consequences and the rheological properties of the suspensions is dominated by hydrodynamic, contact and colloidal interactions between the particles.

In the simplest situation where only Coulombian contact and hydrodynamic interactions are taken into account, and when inertia is negligible, the stress is linear with respect to the shear rate \citep{Boyer2011}. However, even in simple shear, the normal components of the stress are non-zero \citep{Dbouk} and it can cause a migration of one phase relative to the other. In addition, the normal stresses strongly depends on volume fraction\citep{guazzelli_pouliquen_2018}.
One of the difficulty encountered in practical situations originates from the strong coupling between the flow field and the field of volume fraction. Indeed, particles tend to migrate towards low-shear regions of the flow, e.g. the centerline in Poiseuille flow \citep{karnis,lyon1998} or the outer cyldinder in Couette geoemetry \citep{Boyer2011b,sarabian2019}. Conversely, homogeneous shear tends to disperse particles analogously to diffusion \citep{Leighton1986}. This phenomenon, referred to as \textit{shear-induced migration} in the literature, is responsible for viscous resuspension. When a non-brownian suspension of buoyant particles is sheared, migration competes with sedimentation \citep{Leighton1986,Acrivos1993}.   

Shear induced migration has been first modeled using a an apparent diffusion coefficient which is proportional to the shear rate and which strongly depends on the volume fraction $\phi$ \citep{Leighton1986}. Refinements of this diffusive approach have taken into account the fact that it is anisotropic \citep{phillips1992} and that it has several origins: gradients of shear rate, of volume fraction and of viscosity. Another type of theoretical approach has emerged and is based on a momentum balance in the particle phase \citep{Morris1999}. In addition to the fact that the so-called suspension balance model is tensorial and thus accounts for the anisotropy of the migration, its main interest stems from the fact that the particle stress tensor, which is responsible for particle migration can be determined independently \citep{Zarraga2000,Boyer2011b}, using measurements of the normal components of the stress. Despite its relative success to account for the experimental data available, a discrepancy exists in the literature concerning the volume fraction dependency of the particle normal stress components \citep{SaintMichel}. This disagreement concerns both the dilute regime where the particle normal stress scales either as $\phi^2$ \citep{Morris1999,Boyer2011b} either as $\phi^{3}$ \citep{Zarraga2000} and both the divergence exponent when approaching a maximum volume fraction $\phi_m$. Direct measurements of particle migration in Poiseuille flow failed to resolve this discrepancy \citep{guazzelli2015}, as the fully developed volume fraction profile is only weakly sensitive to the exact form of the function used to model the normal stress. However, two series of experiments aimed at revisiting viscous resuspension \citep{Dambrosio,SaintMichel} have been able to test the models with a good accuracy. They took advantage of the steady state profile of volume fraction when shearing a suspension of buoyant particles in a Couette cell. Unfortunately, the two sets of results lead to different conclusions. In \citep{SaintMichel}, it is found that Boyer's correlation \citep{Boyer2011b} was accounting for the experimental data, whereas in \citep{Dambrosio}, Zarraga's correlation \citep{Zarraga2000} better matches the measurements. In addition to this disagreement, a non-linear relation between the stress and the shear rate was found in both cases, suggesting that the systems under study does not correspond to the simple case of Coulombian contact. Thus, the exact form of the volume fraction dependency of the particle normal stress responsible for particle migration is still an open question, which represent a strong limitation of the use of the suspension balance model to predict the volume fraction and flow fields of a sheared suspension. 

Yet, all the available descriptions of the particle stress tensor suppose that all the normal components of the particle stress tensor share the same volume fraction dependency and only differ by a numerical prefactor \citep{Zarraga2000,guazzelli_pouliquen_2018}, although there has been experimental and numerical indications \citep{Dbouk,Gallier2016} that reality is not as simple. It might also be the reason why several empirical laws have been proposed.  Thus, it is needed to determine accurately the particle stress in all the relevant directions. Due to its high sensitivity and its ability to cover a large range of volume fraction, viscous resuspension appears to be a relevant choice for a reference case addressing this issue. From an experimental point of view, it has been used in both cases where gravity is aligned with the vorticity direction (Couette geometry \citep{Acrivos1993,SaintMichel,Dambrosio}) and with the shear direction \citep{Leighton1986}. It has also been tested in pipe flows \citep{norman2005}, but the complexity of the flow field in this case hinders accurate determinations of the particle stress tensor. 


In this work particle stresses are analyzed numerically, using a model which combines solid contact and lubrication in a visco-elastic pair-interaction model~\cite{chevremont2019,chevremont2020}. Here lubrication refers to the terms which, at small gap, dominates the viscous resistance to relative motion between immersed particle pairs. The model also accounts for surface roughness. Homogeneous shear simulated on this basis \cite{chevremont2019}, while neglecting all hydrodynamic contributions beyond lubrication, were in quantitative agreement with available experimental data down to volume fractions of about 20\%. In the present work, not only we extended the results of \cite{chevremont2019} to the three normal stress components using homogeneous shear simulations, but also we determined them indirectly (in the shear and in the vorticity direction) by simulating viscous resuspension, similarly to recently used experimental protocols \cite{SaintMichel,Dambrosio}. Combining the two configurations had several interest, in addition to verify the consistency of the approach. Viscous resuspension allows to determine in a single simulation a whole range of volume fractions, and is thus highly efficient, but cannot be used to access very low volume fractions. It also suffers from two main drawbacks which should be clearly taken into account in experimental setups. The first one is linked to the strong concentration gradients at the interface between the particle bed and the suspending liquid. We show here that they prevent a continuum description at low volume fractions. The second one is related to inertial effects, which we quantify accurately, and which are found to affect data from the literature.


The paper is organized as follow. We recall briefly in section \ref{sec:material_methods} the numerical method used and the SBM framework. In section \ref{sec:results}, we first detail the normal stress results determined in simple shear experiments of homogeneous suspensions of non-buoyant spheres. Second, the resuspension results are presented, analyzed and compared to the homogeneous volume fraction case and to available experiments.

\section{\label{sec:material_methods}Model and methods}

\subsection{\label{ssec:EoM}Equations of motion and interactions}

The motion of suspended particles is integrated in time using the Discrete Element Method (DEM) implemented in Yade-DEM \citep{yade:background2}. 
The method is based on an explicit time-integration of the Newton's equations of motion, for each particle:
\begin{equation}
\frac{d}{dt}\left(\begin{array}{c}
m\dot{\mathbf r}\\
\mathbf{ J\Omega}
\end{array}\right) = \sum \left(\begin{array}{c}
\mathbf F\\
\mathbf T
\end{array}\right)
\end{equation}
where $\mathbf r$ is the position of the center of mass of the particle and $\dot{\mathbf r}$ its time derivative, $m$ is the mass, $\mathbf J$ is the moment of inertia tensor, $\mathbf \Omega$ is the rotational velocity vector, $\mathbf F$ denotes forces acting on the body and $\mathbf T$ the moments of these forces about $\mathbf r$.

\begin{figure}
    \centering
    \includegraphics[width=0.6\linewidth]{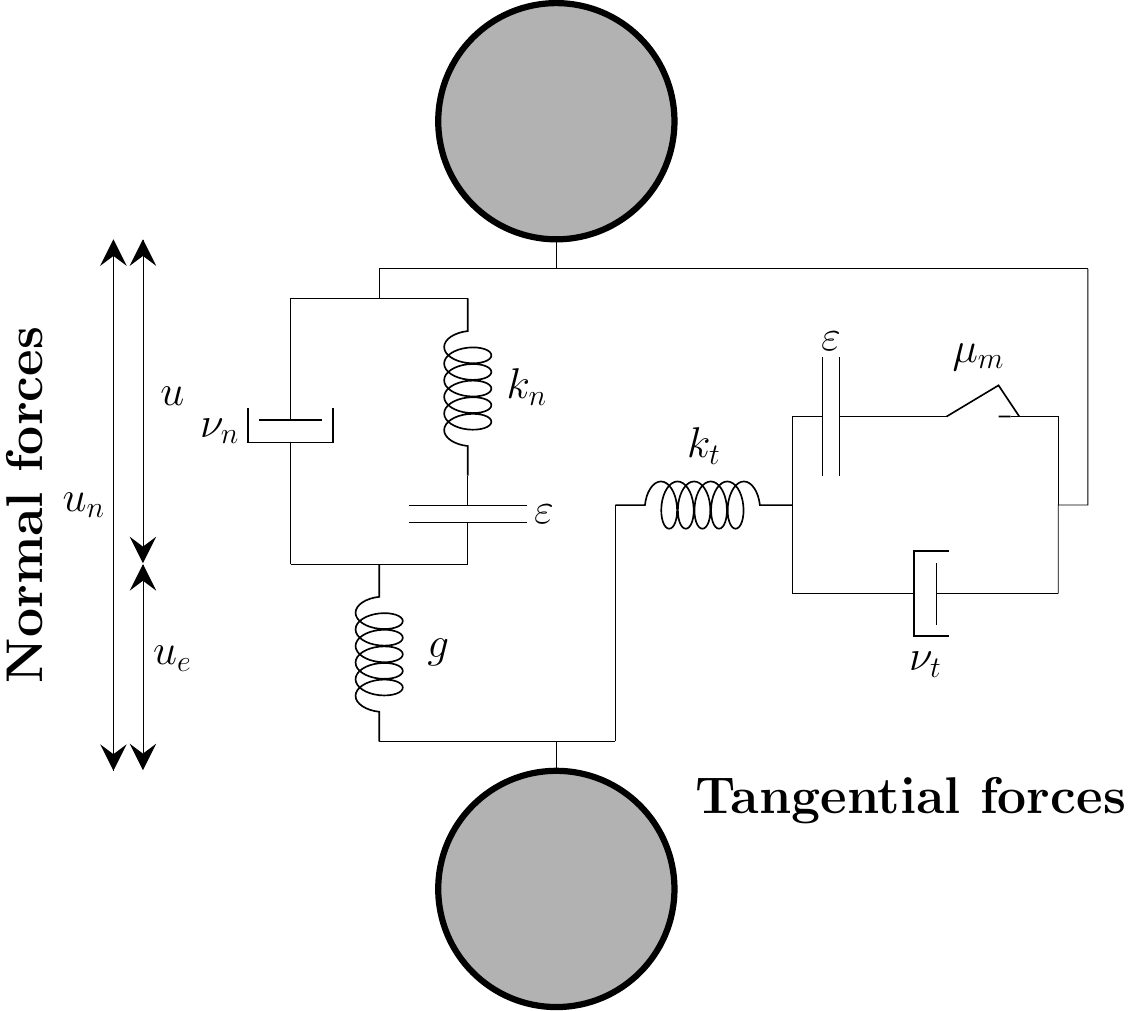}
    \caption{Rheological scheme of the lubricated contact model developed in \cite{chevremont2020}. $u_n$ is the geometrical distance between undeformed spheres surfaces, $u_e$ is surface deformation because of the interaction, $u$ is the gap between deformed surfaces. $\nu_n$ is the normal viscosity coefficient from lubrication, $\varepsilon$ is the dimensionless roughness, $k_n$ and $g$ are the roughness and bulk stiffness, here with equal values. $\nu_t$ is the tangential viscosity coefficient, $\mu_m$ the friction coefficient and $k_t$ the tangential stiffness of roughness.}
    \label{fig:lubricated_contact_model}
\end{figure}

Figure \ref{fig:lubricated_contact_model} summarizes the rheology of the lubricated contact model developed in \cite{chevremont2020} for slightly deformable particles with rough surfaces. The model couples frictional contacts and lubrication forces in a single visco-elasto-plastic model, as follows. 

First, let $u_n$ the change of center-to-center distance with respect to the force-free configuration for particles of radius $a$: $u_n = |\bm r_1 - \bm r_2| - 2a$, with $\bm r_{1,2}$ the position vectors. $a$ is the pair average particle radius ($a=(r_1+r_2)/2$.
If $u_n=0$ and the particles do not move ($\dot u_n=0$), there is no interaction force. It is assumed that in this situation a liquid film remains in a gap of thickness $\epsilon a$ between the solid surfaces, where the coefficient $\epsilon$ defines the length of asperities at the surface of the particles.
It is further assumed that any change of the gap between the surfaces will cause an interaction force between the particles which is, in the general case, the sum of of two forces. The first is an elastic force originating from the compression of the asperities if the gap between the surfaces is less than $\epsilon a$:
\begin{gather}
\mathbf F^c_{n} = -k_n \max(0, \varepsilon a - u)\mathbf{n}.
\end{gather}
Here, $u$ denotes the gap, $k_n$ is the stiffness of the asperities, and $\mathbf{n}$ is the unit normal of the contact plane.
The other force comes from lubrication by the suspending liquid and it is proportional to the rate of change of the gap, following \cite{FrankelAcrivos1967}:
\begin{gather}
  \mathbf F_n^l = \nu_na^2\dfrac{\dot u}{u}\mathbf{n},
\end{gather}
where $\nu$ is the viscosity of the liquid. 
The contact and lubrication force, together, cause a small deflection of the surfaces, $u_e$, proportional to the total normal force. Note that this coupling method is not a cut-off of the lubrication term. Contact and lubrication exists together, as showed in \cite{chevremont2020}. TODO: reformuler pour mettre de facon positive. We note the deflection $u_e$, and we assume a linear elastic response such that 
\begin{gather}
 \bm F_n^l + \bm F_n^c   = g u_e\bm n, \label{eq:normalDeflextion}
\end{gather}
where $k_b$ is the stiffness of the particles. Hereafter, we assume $g = k_n$ for simplicity.

With the above definitions the displacements satisfy the equality $u_n = u + u_e$: any change of the center-to-center distance cause a change of the gap distance, or a deformation of the particles (or both). Combining the above equations leads to a differential equation for $u$ which is integrated in time to track the evolution of the normal interaction force. The details of this integration can be found in \cite{chevremont2020}. 

In addition, a shear force results from tangential velocity $\dot{ \bm v}$ between the particles. It is also the sum of two terms. The first term comes from the solid contact, it increases linearly with $\bm v$, with a stiffness $k_t$, until it reaches a threshold magnitude defined by Coulombian friction:
\begin{gather}
\lVert \mathbf F^c_s \rVert \le \mu_m\lVert \mathbf F^c_n \rVert, \label{eq:coulomb},
\end{gather}
with $\mu_m$ the friction coefficient.
The other term is a lubrication term, again following \cite{FrankelAcrivos1967}:
\begin{gather}
\mathbf F_s^l = \dfrac{\pi\eta_f}{2}\left[-2a+(2a+u)\ln\left(\dfrac{2a+u}{u}\right)\right]\mathbf{\dot v}
\end{gather}

The force terms associated to contact (both normal and tangential) vanish as soon as the gap is larger than $\epsilon a$. In such case only the lubrication terms remain. They decay rapidly with distance. A cut-off distance equal to $2a$ is selected, no lubrication is computed beyond that. For all simulations in this work, the stiffness is high enough to stay in the rigid-rough case. When the particles interact from a distance by lubrication alone, the model almost reduces to a visco-elastic variant of the conventional lubrication models.

\subsection{\label{ssec:simpleShear}Simple shear simulations}

Homogeneous simple shear simulations are carried out with tri-periodic boundary conditions at imposed volume fractions from $\phi=0.35$ to $\phi=0.995\phi_m$ with $N=10000$ neutrally buoyant spheres, with $\phi_m$ the volume fraction of divergence. The sphere radii are distributed around $a$ using a Gaussian distribution with a standard deviation of $5\%$. This small polydispersity prevents the system from crystallizing. The relative roughness is set to $\varepsilon=10^{-3}$.  The friction coefficient is varied from $0.1$ to $0.5$, each value resulting in a different $\phi_m$ \citep{chevremont2019}. 

\begin{figure}
    \centering
    \subfloat[Simple shear cell\label{fig:simple_shear_cell}]{\includegraphics[width=0.33\linewidth]{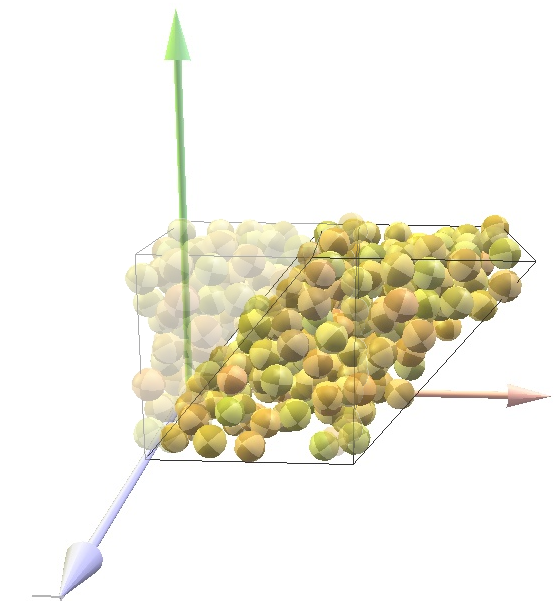}}
    \subfloat[Viscous resuspension in the gradient direction\label{fig:resuspension22_cell}]{\includegraphics[width=0.33\linewidth]{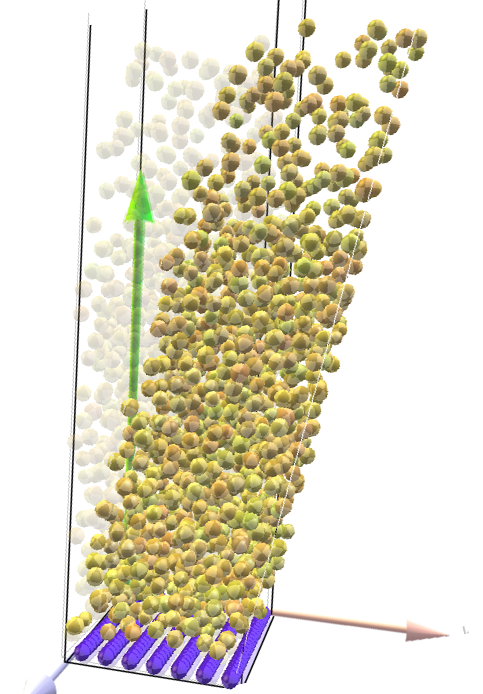}}
    \subfloat[Viscous resuspension in the vorticity direction\label{fig:resuspension33_cell}]{\includegraphics[width=0.25\linewidth]{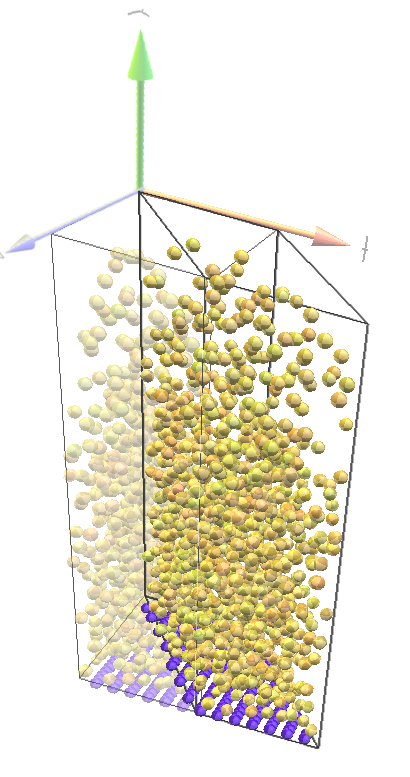}}
    \caption{Snapshots of simulation cell for the tree types of simulations. For resuspension cases gravity is oriented against the 2-axis (green arrow). The dark blue spheres are forming a floor for the suspended (yellow) spheres.}
    \label{fig:cell_snapshot}
\end{figure}

The initial state for the shear flow is obtained by compressing a very dilute suspension made of $N$ spheres with random initial positions, until the desired $\phi$ is reached.
The periodic cell is then imposed a homogeneous velocity gradient $\dot\gamma \bm e_1\otimes \bm e_2$ (Fig. \ref{fig:simple_shear_cell} and the stress components are evaluated in the steady state. 

\subsection{\label{ssec:viscous resuspension}Viscous resuspension simulations}

Viscous resuspension is simulated with gravity along the 2-axis, with $\bm g$ = (0,-g,0), and periodic boundaries in the other directions. The entire system is supported by a floor made of array of particles which remain at a fixed height (the blue ones in Fig. \ref{fig:cell_snapshot}). We investigate two types of deformation: vorticity and gravity are othogonal in the first case, they are colinear in the other case (FIXME: fig).
The simulations comprise $N=10000$ spheres, with the same particle size distribution and the same interaction parameters as for simple shear. Friction coefficient is set to $\mu_m=0.5$, so that $\phi_m=0.58$.
Figure \ref{fig:resuspension22_cell} shows the simulation cell when resuspension occurs in the gradient direction.
For this specific case, external drag forces are introduced to force homogeneous shear at the macroscale. The external drag is proportional to the deviation between particle velocity and mean-field velocity, assuming Stoke's drag with a viscosity equal to $1/30$ times the viscosity considered for lubrication.
It should be emphasized that this setup is not equivalent to conditions found in most gravitational flows (bedload, typically). Indeed most experimental conditions lead to homogeneous - or linear - shear stress instead of homogeneous shear rate.

Figure \ref{fig:resuspension33_cell} shows the simulation cell when the resuspension occurs in the vorticity direction. In this case the motion of the particles forming the floor follows the macroscopic velocity gradient.

The inertial effects in the simulations are negligibly small. They can be quantified by the particles Reynolds number $\mathcal Re_p=\rho \dot\gamma a^2/\eta_f$. This number stays below $5~10^{-3}$ for all simulations - with an exception for some series specifically dedicated to analyzing the role of inertia in section \ref{subsec:inertia}.

\subsection{\label{ssec:simpleShear}Particle Phase Stress}
In homogeneous conditions (section \ref{ssec:simpleShear}), the \emph{virial stress} gives an unambiguous definition of the mechanical stress in the system of pair-interacting particles. The virial stress (FIXME: ref) reads: $\bm \sigma = <\bm f\otimes\bm l> - m<\bm v \otimes \bm v>$. The first term on the right-hand side is the volume-averaged outer product of the pairwise interaction forces $\bm f$ (from both lubrication and direct contact) and the branch vector $\bm l$ connecting reference points associated to the particles. The second term involves inertial effects associated to the velocity fluctuations $\bm v$, it is negligibly small by assumption in the viscous regime. In our simulations we stay in the viscous, non-inertial, regime. (FIXME: quantifier et signaler l'exception). This expression is used to evaluate stresses in simple shear, consistently with [FIXME Lhuillier, SBM revisited].

Inhomogeneous, buoyant, conditions (section \ref{ssec:viscous resuspension}) enable an alternative evaluation of some stress components at steady state. Indeed, the concentration profile should satisfy the momentum balance in the particle phase, which reads~\citep{Morris1999}: 
\begin{equation}
    \phi \Delta \rho \mathbf{g} + \mathbf{\nabla} \cdot {\bm \sigma} =0\,,
    \label{eq:sbmbase}
\end{equation}
where $\bm \sigma$ is the particle phase stress tensor, of diagonal components  $\sigma_{ii}=-\eta_0 \left|\dot{\gamma}\right| \eta_{ii}\left(\phi\right), i=1, 2, 3$. They involves the normal viscosity coefficients $\eta_{22}$ and $\eta_{33}$, respectively. The above momentum balance leads to
\begin{equation}
   \frac{\phi}{{\Sh} } = -\frac{\rm d \eta_{ii}}{\rm d \phi} \frac{{\rm d} \phi}{{\rm d} {x_i}}\,,
    \label{eq:sbmdimensionless}
\end{equation}
where $i=2$ or 3, $x_i$ is the coordinate normalized by the particle size $a$ and where we have introduced the Shields number $\Sh$, defined as 
\begin{equation}
    {\Sh} = \frac{\eta_0 \dot\gamma}{\Delta \rho g a}.
    \label{eq:shields}
\end{equation}

The volume fraction profile can be used to determine the normal viscosity directly. Equation \ref{eq:sbmdimensionless} can be integrated directly from any position $y$ along the resuspended height, which leads to:
\begin{equation}
    \eta_{ii} = \frac{1}{\Sh}\int_{\hat z}^{+\infty}\phi(u)du \label{eq:eta_from_phi}
\end{equation}
Integrating to the infinity or to resuspended height is the same, as the volume fraction is $0$ above the resuspended height. This expression is more robust, as it doesn't require to determine the resuspended height first.

\subsubsection{FIXME:particle phase stress in the litterature}
Several temptative empirical expressions for this quantity --also called correlations in the literature~\citep{Acrivos1993,Guazzelli2018}-- have been proposed, which generally assume the following form:
\begin{equation}
\label{eq:normvisc_correlations}
    \eta_{ii}(\phi) = \lambda_{i} \left ( \frac{\phi / \phi_m}{1-\phi/\phi_m} \right )^n\,,
\end{equation}
where $\phi_m$ is the volume fraction at which both shear and normal viscosities diverge. \citep{Zarraga2000} choose $n = 3$, $\lambda_{1}=1.5$, $\lambda_{2}=1.36$, $\lambda_{3} = 0.62$ 
and $\phi_m = 0.62$ based on previous experimental results of viscous resuspension in a Taylor-Couette geometry~\citep{Acrivos1993}. \citep{Morris1999} obtain $n = 2$, $\lambda_{3}=0.38$ and $\phi_m=0.68$ combining sets of data from large-gap Taylor-Couette and parallel-plate migration experiments. This scaling is very similar to the one proposed by \citeauthor{Boyer2011b}~\citep{Boyer2011b,Boyer2011} who derived  $n=2$, $\lambda_{3} = 0.6$ and $\phi_m=0.585$ from pressure-imposed shear and rotating rod experiments.

\section{\label{sec:results}Results and discussion}

\subsection{\label{ssec:simpleshear}Simple shear}

\begin{figure}
    \centering
    \includegraphics[width=0.7\linewidth]{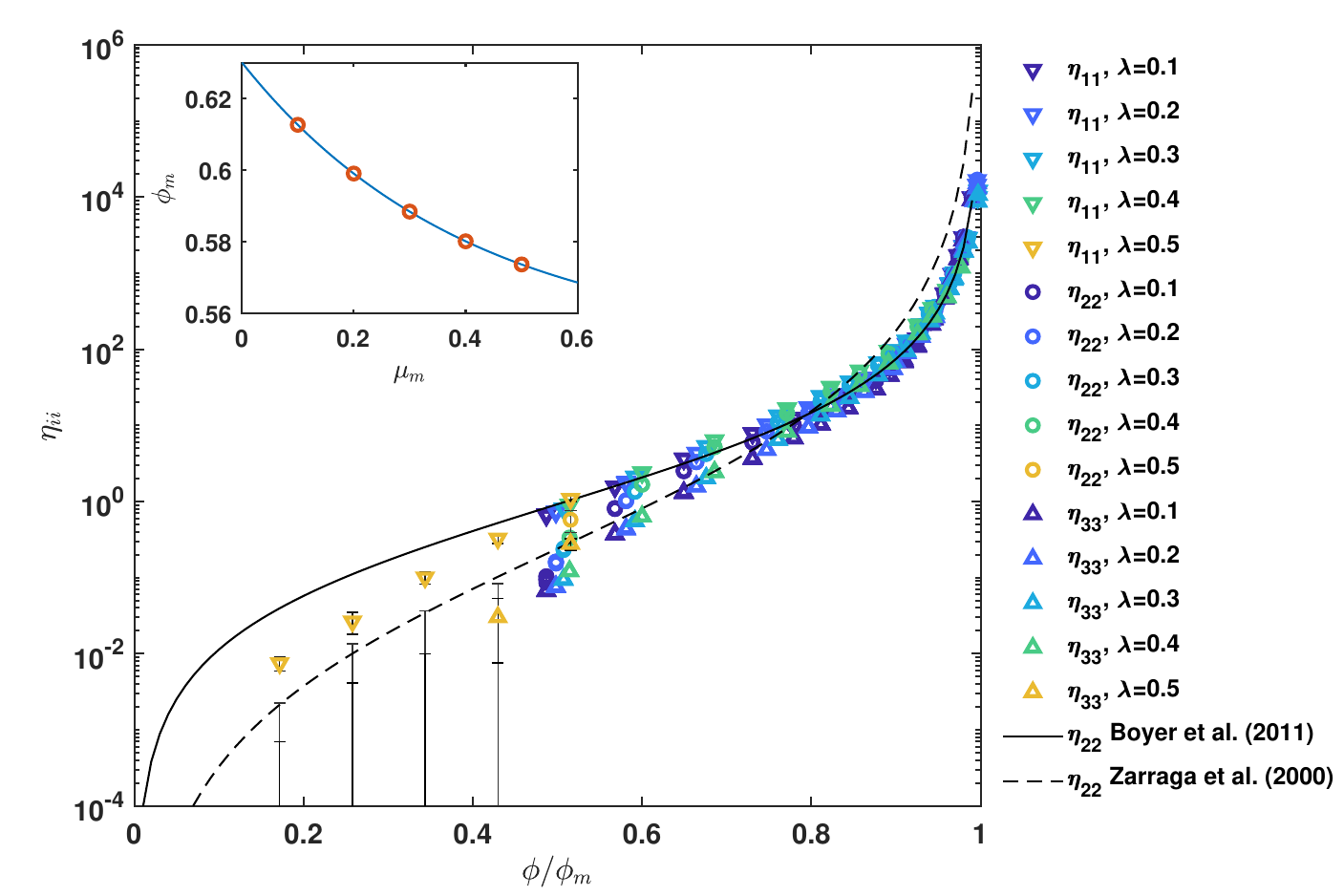}
    \caption{Normal viscosities from simple shear experiments as a function of $\phi/\phi_m$. The error bars indicate the standard deviation on $\eta_{22}$ and $\eta_{33}$ when they become negligibly small in average. The solid line is the expression from \citep{Boyer2011b}, the dotted line is expression from \citep{Zarraga2000}. The insert shows the dependency of $\phi_m$ on contact friction coefficient $\mu_m$ in the results, with the expression from \citep{chevremont2019} superimposed.}
    \label{fig:normal_viscosities_phi_phim}
\end{figure}

Figure \ref{fig:normal_viscosities_phi_phim} shows the normal viscosities in the three directions as a function of $\phi/\phi_m$ and for various microscopic friction coefficents $\mu_m$. $\mu_m$ has a direct consequence on the maximal volume fraction $\phi_m$ \citep{chevremont2019}. Nevertheless, plotting versus $\phi/\phi_m$ as in Figure \ref{fig:normal_viscosities_phi_phim} suggests an approximate collapse on a single curve independently of $\mu_m$. When plotted versus $\phi$ (not shown), such a collapse is not obtained. Note that the precise relation between $\phi_m$ and $\mu_m$ is likely to be model dependent. 

For $\phi/\phi_m > 0.5$, the viscosity coefficients are ordered like $\eta_{11} > \eta_{22} > \eta_{33}$ with $\eta_{22}$ and $\eta_{33}$ close to each other. On the one hand, all viscosities are approximately equal close when approaching the maximum volume fraction, and are well captured by \citep{Boyer2011b}'s expression, with a $-2$ divergence. On the other hand, for more dilute suspensions, \citep{Zarraga2000}'s expression accounts more closely for the data of $\eta_{22}$ and $\eta_{33}$. In the more dilute regime, $\eta_{11}$ still follows the same trend as \citep{Boyer2011b}'s expression, whereas $\eta_{22}$ and $\eta_{33}$ drops quickly to zero. In that case, they are represented by their standard deviation, which is much higher than the mean value. The volume fractions at which $\eta_{22}$ and $\eta_{33}$ vanish are not the same, as found by other authors \citep{Gallier2014}.

\begin{figure}
    \centering
    \includegraphics[width=0.7\linewidth]{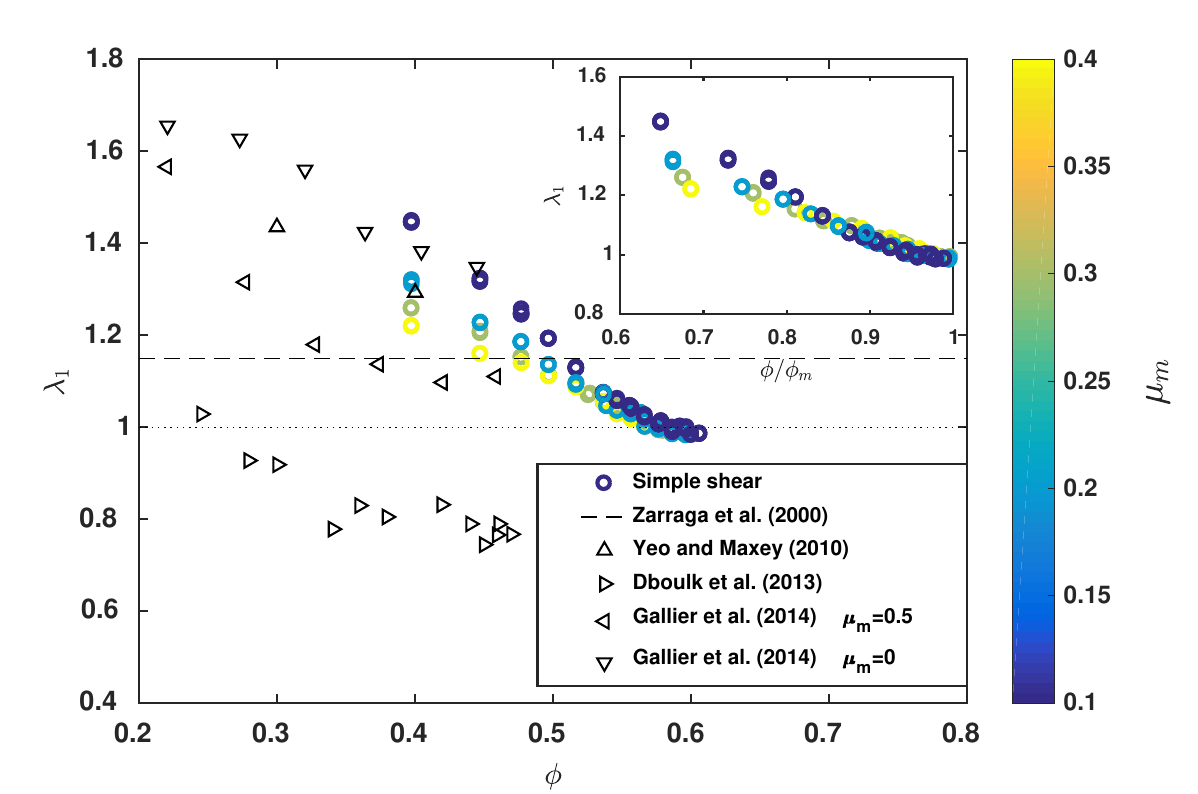}
    \caption{$\lambda_{1}=\sigma_{11}/\sigma_{22}$ as a function of $\phi$ and with various friction coefficients, from simple shear simulations and from literature \citep{Zarraga2000,yeo2010,Dbouk,Gallier2014}}
    \label{fig:lambda1}
\end{figure}

In the following, all stresses are rescaled by $\sigma_{22}$, which is experimentally measurable and controllable. The ratio $\lambda_{1}=\sigma_{11}/\sigma_{22}$ from our results in simple shear and from the literature is plotted on figure \ref{fig:lambda1}. Note that rescaling $\phi$ by $\phi_m$ in that plot does not lead to a single master curve, as shown in the insert. Indeed, this ratio depends on the microscopic friction coefficient. Data from \citep{Gallier2014} and \citep{yeo2010} and relative to two different friction coefficients ($0$ and $0.5$ respectively) are consistent with the present ones. The value of $\lambda_{1}$ proposed in \citep{Zarraga2000} is also in the range of our data. On the other hand, the points reflecting measurements by \citep{Dbouk} are outliers (all other points are from numerical simulations).

\begin{figure}
    \centering
    \includegraphics[width=0.7\linewidth]{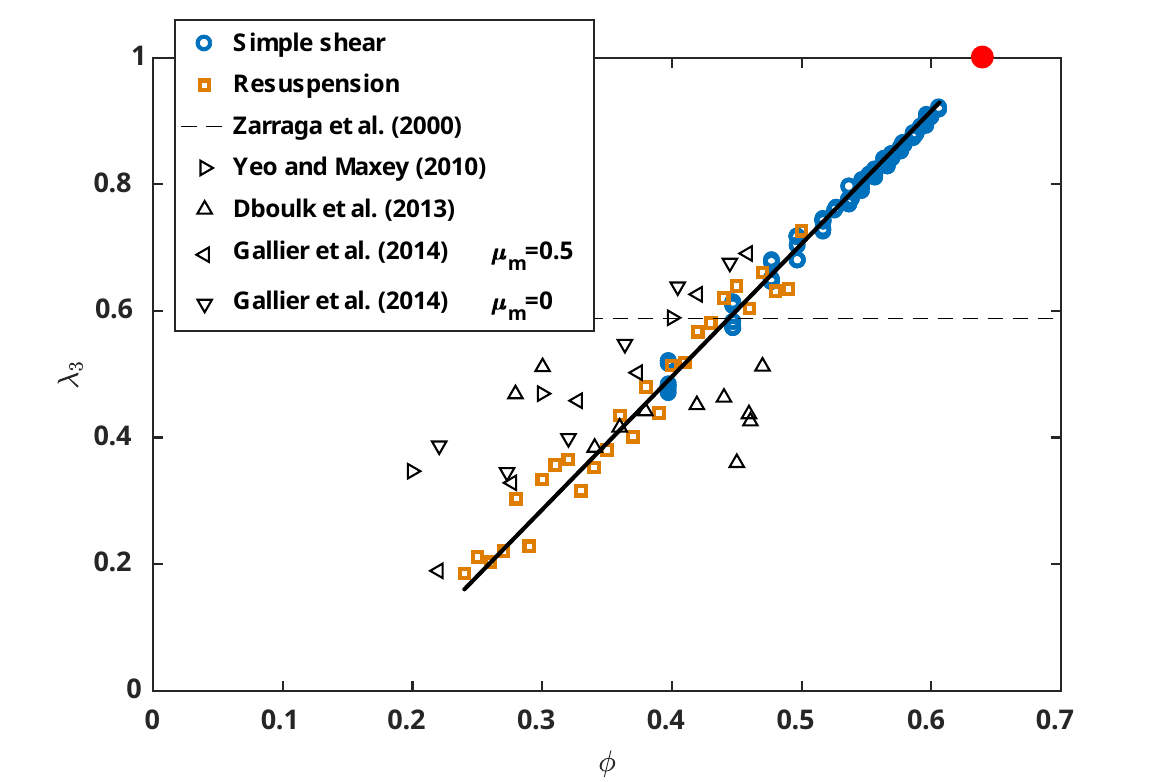}
    \caption{$\lambda_{3}=\sigma_{33}/\sigma_{22}$ as a function of $\phi$, from simple shear, resuspension and from literature \citep{Zarraga2000,yeo2010,Dbouk,Gallier2014}. Red point is at coordinate $(0.64;1)$. Solid line: best linear fit.}
    \label{fig:lambda3}
\end{figure}

Figure \ref{fig:lambda3} shows $\lambda_{3}$, the ratio between $\sigma_{33}$ and $\sigma_{22}$ as a function of $\phi$, based on literature data, simple shear and resuspension simulations (the latter being detailed in the next section). In contrast with $\lambda_{1}(\phi)$, the $\lambda_{3}(\phi)$ results collapses to a single straight line. It means that $\lambda_{3}$ is independent on the microscopic friction coefficient, and thus on $\phi_m$. The present data are also close to the data from the literature, but much less dispersed. A linear fit leads to 
\begin{align}
    \lambda_{3}(\phi) & = 2.1(\phi-\phi_{rcp})+1  &\text{   if   } 0.1638 < \phi < 0.64\label{eq:lambda33lin}  \\
     & = 0  &\text{  if  } \phi \leq 0.1638
\end{align}
with $\phi_{rcp}=0.64$, the random close packing volume fraction. That $\lambda_{3}$ vanishes at low $\phi$ instead of following the linear trend in the negative domain is a conjecture, based on the fact that suspensions are always dispersive. There is in fact no data available in the corresponding range.

It is remarkable that $\lambda_{3}$ tends to $1$ at $\phi_{rcp}$ and that it depends only on $\phi$. In contrast, $\lambda_{1}$ as a complex dependency on contact friction through $\mu_m$. This expression vanishes at low volume fraction, as it should remain positive. For practical use in continuum mechanics, it might be more convenient to define a continuous and derivable function. However, our data are not sufficient to describe how the function tends toward $0$. 

Another way of representing the data is to compute the first and second normal stress differences, $N_1 = \sigma_{11}-\sigma_{22}$ and $N_2 = \sigma_{11}-\sigma_{22}$, and the particle pressure $\Pi = \sigma/3$. A plot of these quantities is available as a supplemental material

\begin{figure}
    \centering
    \includegraphics[width=0.7\linewidth]{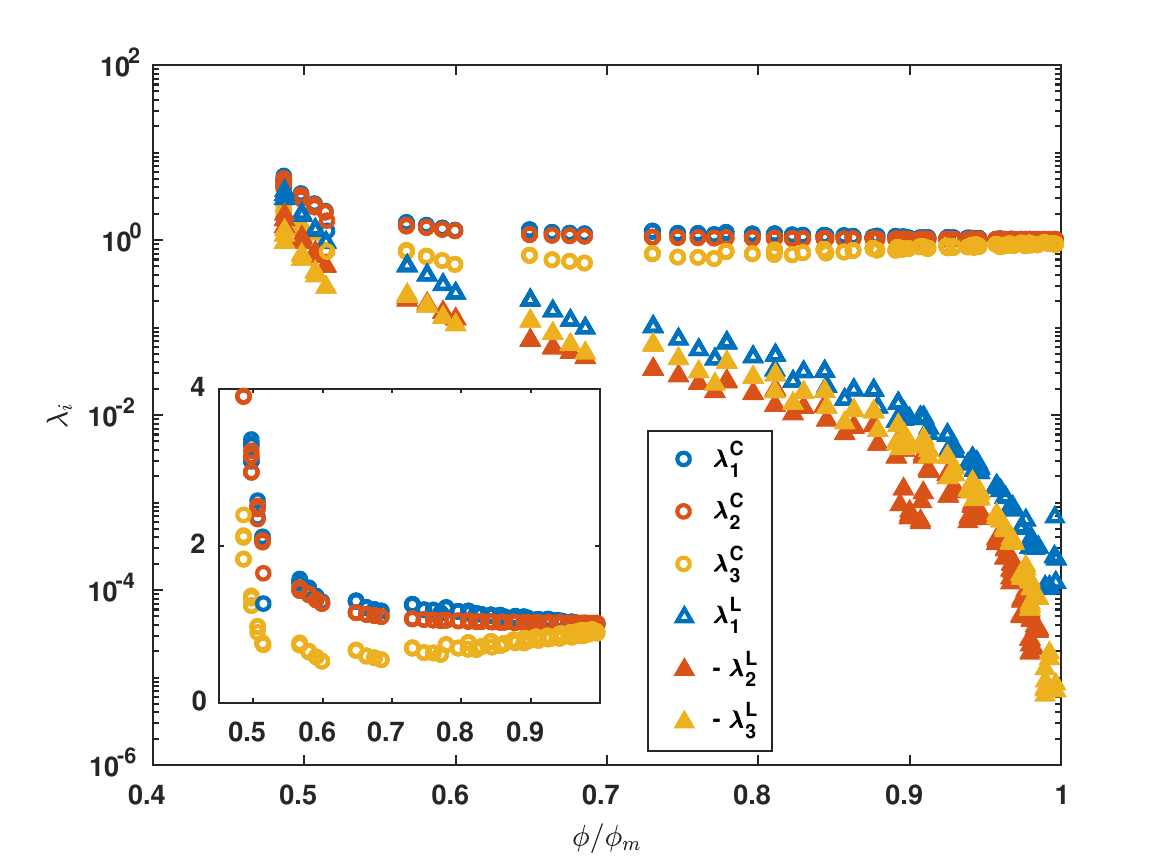}
    \caption{Ratio $\lambda_{i}^k = \sigma_{ii}^k/\sigma_{22}$, where $k$ denotes $L$ubrication or $C$ontact. Open symbols are positive whereas filled symbols are negative. The insert shows only the contribution from solid contacts}
    \label{fig:lambda_xxyyzz}
\end{figure}

Let us eventually focus on the normal stress decomposition in terms of contact and lubrication.  Figure \ref{fig:lambda_xxyyzz} shows the contact and the lubrication forces contributions to each normal stress components. As previously, the stresses are normalized by $\sigma_{22}$. In this data set, contacts dominate all normal viscosity components, down to $\phi/\phi_m\simeq 0.5$. For the contact contribution, $\lambda_{1}^C$ and $\lambda_{2}^C$ whereas $\lambda_{3}^C$ is smaller. For the lubrication part, we find opposite signs between $\lambda_{1}^L$ which is positive, and $\lambda_{2}^L$ and $\lambda_{3}^L$ which are negative. In the tested range of volume fraction the contribution of lubrication forces to the normal stress remains smaller than that of the contact forces but they are of the same order of magnitude on the dilute side of the range. The fact that the normal viscosities $\eta_{22}$ and $\eta_{33}$ vanishes at low volume fraction could thus be interpreted as a balance of contact and lubrication forces such that their contribution to stress cancel out.      

\subsection{\label{ssec:resuspension} Resuspension}

\begin{figure}
    \centering
    \subfloat[Resuspension in gradient direction\label{fig:phi33_profile}]{\includegraphics[width=0.5\linewidth]{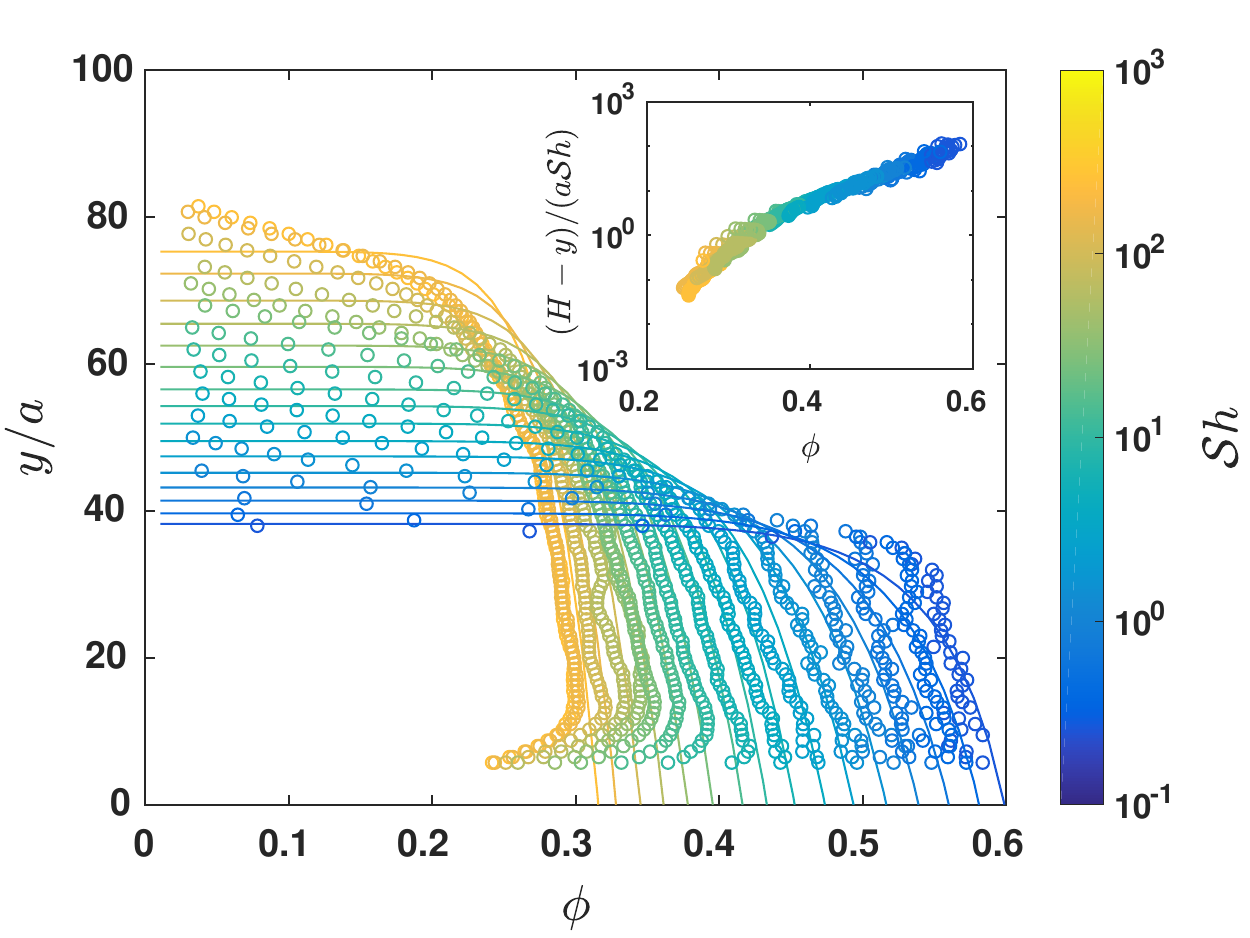}}
    \subfloat[Resuspension in vorticity direction\label{fig:phi22_profile}]{\includegraphics[width=0.5\linewidth]{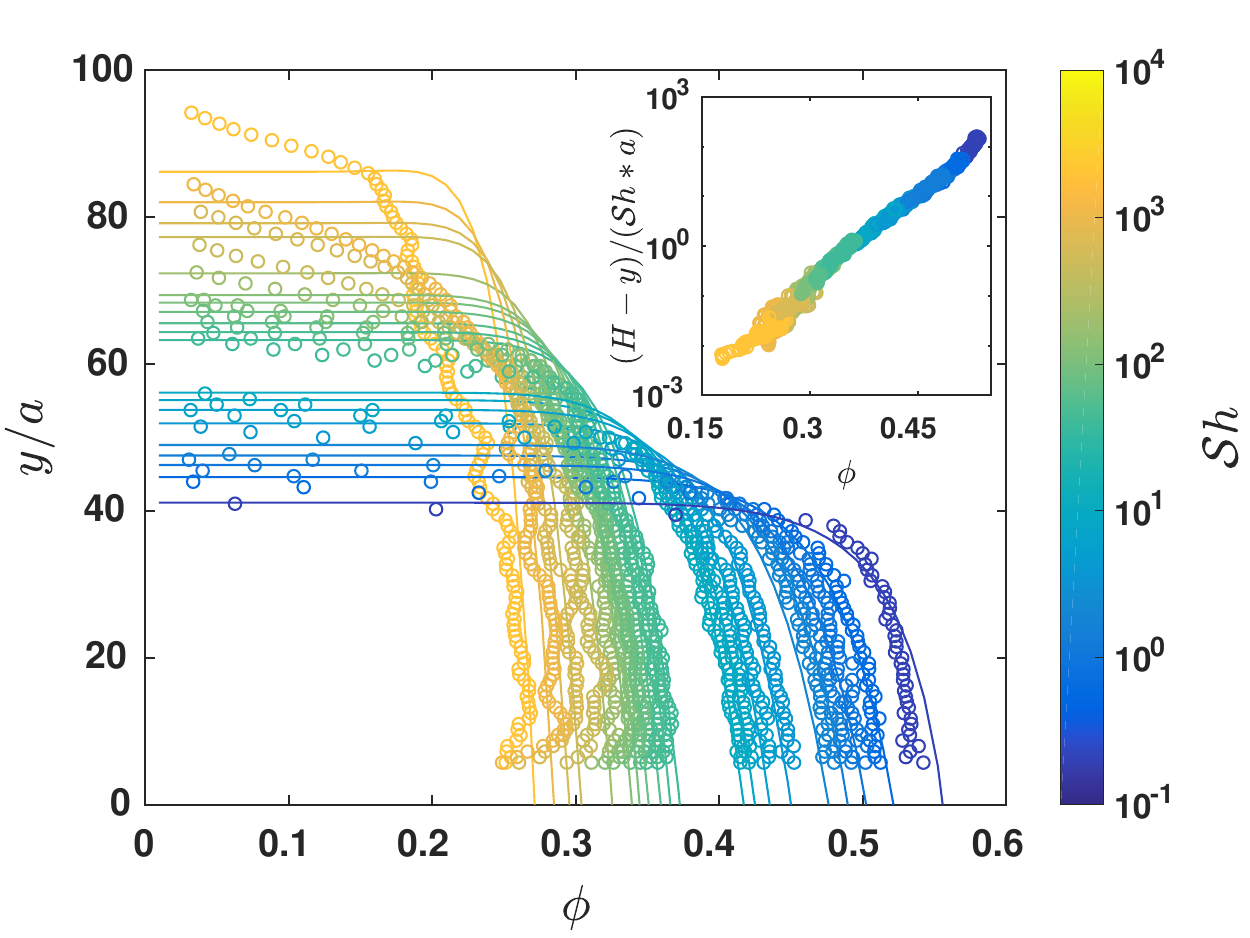}}
    \caption{Raw concentration profiles. Circles are simulation results and solid lines are concentration profiles computed from viscosities empirical expressions. Insert shows the rescaling with Shields number, without the points close to the interface.}\label{fig:phi_profiles}
\end{figure}

Figure \ref{fig:phi_profiles} reflects the steady state concentration profiles obtained in resuspension, in both the shear and vorticity directions. By increasing the Shield number over several orders of magnitude, the profiles evolve from a dense settled bed to a more dilute and inhomogeneous distribution. The volume fraction is typically between 20 and 30\% for the largest Shield number. The gradient of solid fraction is always much larger near the top the suspension. 


\begin{figure}
    \centering
    \subfloat[Resuspension in gradient direction\label{fig:eta33_taken}]{\includegraphics[width=0.5\linewidth]{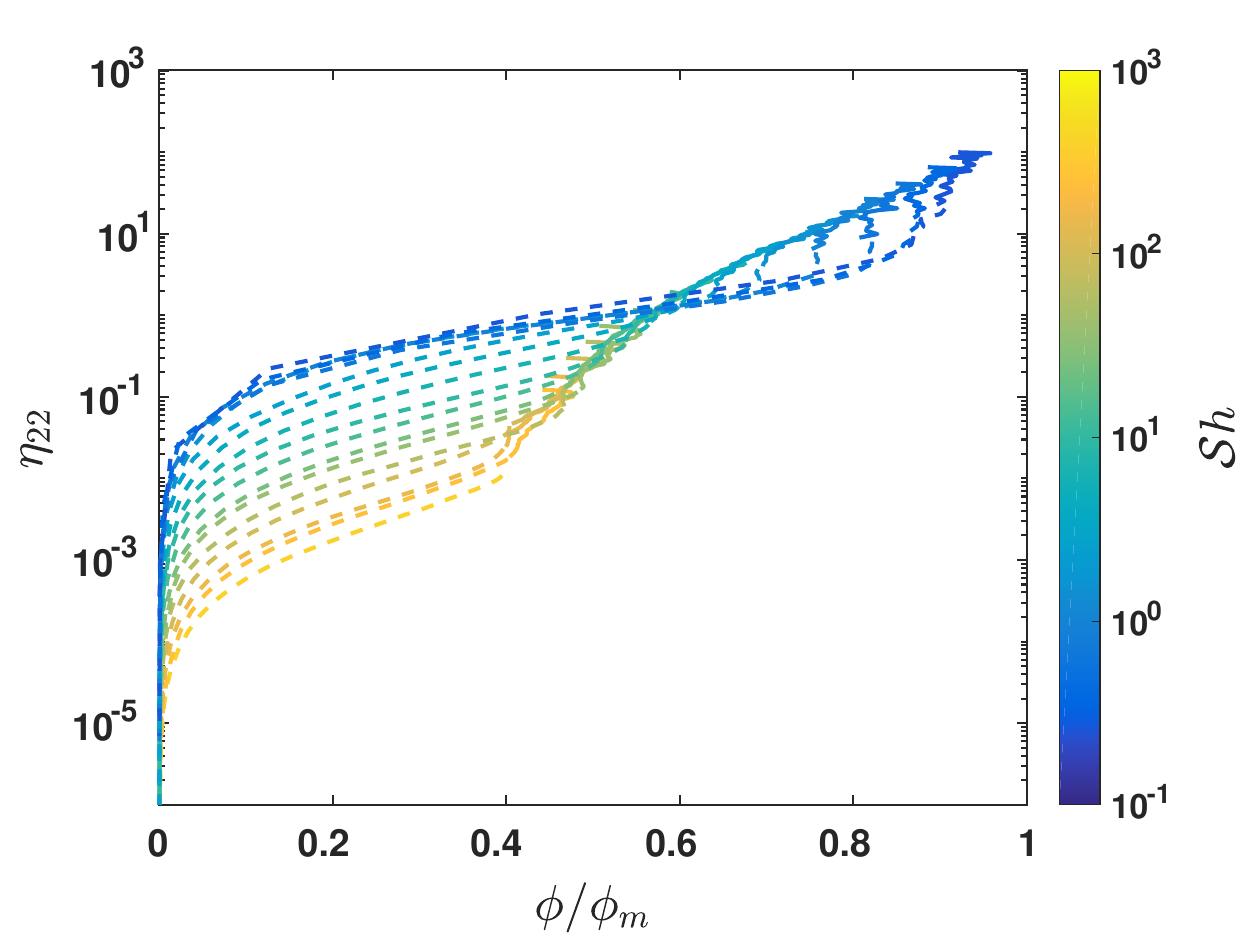}}
    \subfloat[Resuspension in vorticity direction\label{fig:eta22_taken}]{\includegraphics[width=0.5\linewidth]{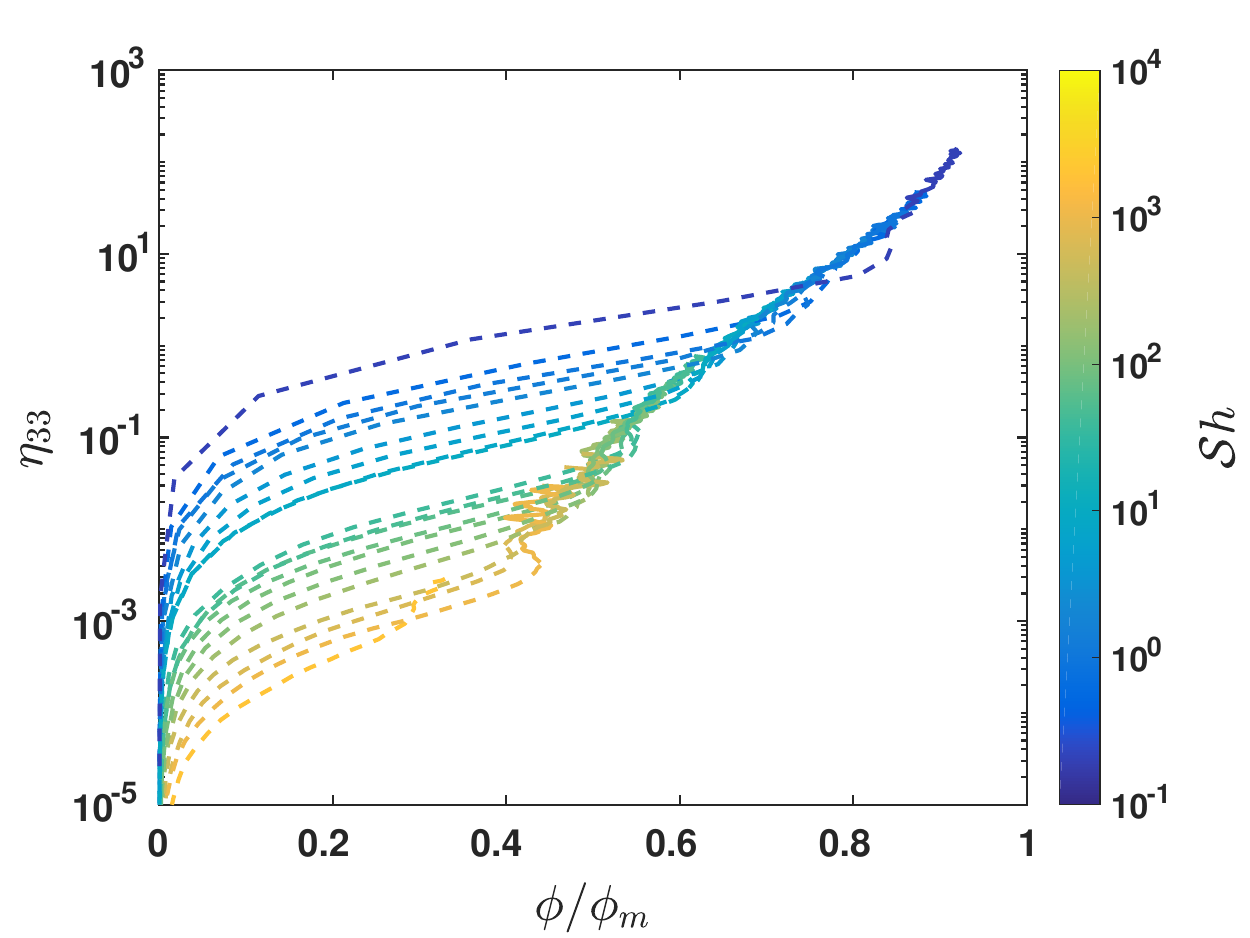}}
    \caption{Raw viscosities as a function of $\phi$, computed from concentration profile using equation \ref{eq:eta_from_phi}. Lines shows the points of the pseudo-plateau which are kept whereas dashed lines shows the points at the interface which are removed.}\label{fig:eta_taken}
\end{figure}

Figure \ref{fig:eta_taken} displays the viscosities $\eta_{22}$ and $\eta_{33}$ computed by integrating all the concentration profiles using equation \ref{eq:eta_from_phi}. On this figure, we can see that part of the data (displayed in solid lines) collapse very well on a master curve, whereas another part (displayed in dotted lines) does not. The latter correspond to the top layers of the re-suspended bed, which exhibit a very strong $\phi-$gradient. Indeed the volume fraction abruptly drops to zero on a distance of $5a$ or even less. It is unsurprising that a homogenized approach cannot explain such an abrupt change in a continuous framework. 

If we discard in the analysis the strong gradient region by ignoring a layer of thickness $6a$ below the interface, the data tends to collapse on a single profile when the length is normalized by the length scale  $a\mathcal{S}h$. The resulting collapsed profiles are displayed in insert in figure \ref{fig:phi_profiles}. 

\begin{figure}
    \centering
    \subfloat[Resuspension in gradient direction\label{fig:eta33_model} ]{\includegraphics[width=0.5\linewidth]{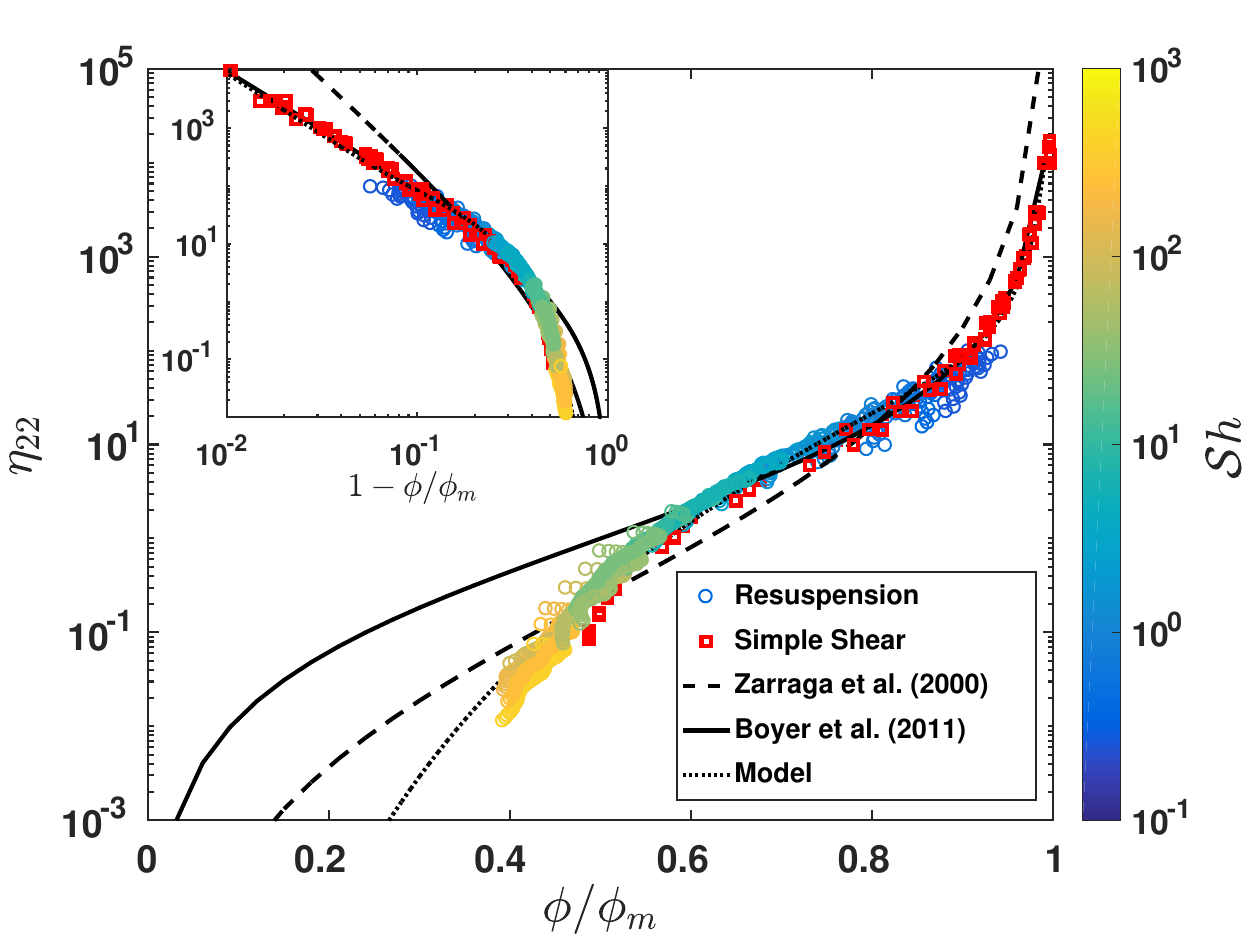}}
    \subfloat[Resuspension in vorticity direction\label{fig:eta22_model} ]{\includegraphics[width=0.5\linewidth]{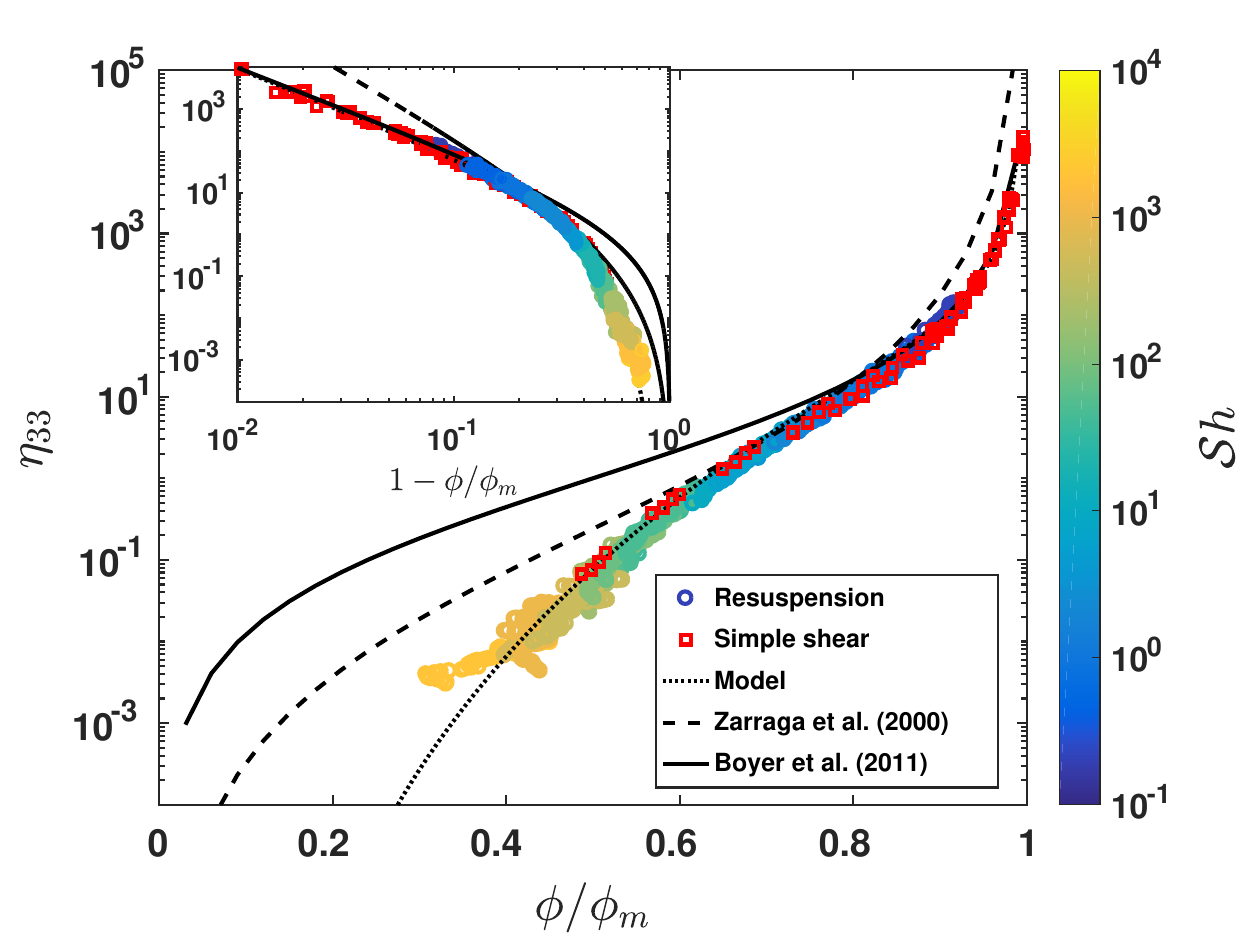}}
    \caption{Viscosity profiles from resuspension and simple shear as a function of $\phi/\phi_m$ with models available in the literature. Inserts shows the data in log-log according to $1-\phi/\phi_m$. Solid line is expression from \citep{Boyer2011b}, dashed line is expression from  \citep{Zarraga2000}. }\label{fig:eta_model}
\end{figure}

Figures \ref{fig:eta_model} shows the computed viscosities as a function of $\phi/\phi_m$, with both data from resuspension and from simple shear. The same data are plotted against $1-\phi/\phi_m$ in the inserts in order to highlight the divergence exponent. A very good agreement is found between the two types of simulations. This clearly shows that the normal stresses indirectly determined in the resuspension tests is the same as the one computed directly in homogeneous shear of a uniform non-buoyant suspension.     

Both $\eta_{22}$ and  $\eta_{33}$ exhibit a $-2$ divergence as the volume fraction approaches the maximal volume fraction, which is well captured by \citep{Boyer2011}'s expression for the normal viscosity. For $\eta_{22}$ (figure \ref{fig:eta33_model}) the expression captures the data rather well down to $\phi/\phi_m\approx0.6$, but it overestimates the coefficient below that value. The same observation can be made on $\eta_{33}$, in that case however \citep{Zarraga2000}'s expression is a much better fit for $0.6<\phi/\phi_m<0.8$. $\eta_{33}$ drops significantly below both empirical expressions as soon as $\phi/\phi_m < 0.6$. 

As none of the earlier proposed expressions can account for the rapid decay of $\eta_{22}$ and $\eta_{33}$ below $\phi/\phi_m=0.6$, we propose in the following equations new expressions that fits the data on the entire range of volume fraction. They read
%

\begin{equation}
 \eta_{22}\left(\phi\right)= \frac{ e^{-3\left(\frac{\phi_m}{\phi} -1 \right)} }{\left(1-\phi/\phi_m\right)^2}+30\,e^{-6\left(\frac{\phi_m}{\phi} -1 \right)}
\end{equation}

\begin{equation}
 \eta_{33}\left(\phi\right)=  \left( 2.1(\phi-\phi_{rcp})+1 \right) \left( \frac{ e^{-3\left(\frac{\phi_m}{\phi} -1 \right)} }{\left(1-\phi/\phi_m\right)^2}+30\,e^{-6\left(\frac{\phi_m}{\phi} -1 \right)} \right)
\end{equation}

The rather complex form of $\eta_{33}$ is coming from the fact that we aimed fits for $\eta_{22}$ and $\lambda_{3}$ (rather than $\eta_{22}$ and $\eta_{33}$), since $\lambda_{3}$ is very sensitive to the particular expressions chosen (see Fig. \ref{fig:lambda3}). The expressions for $\eta_{22}$ and $\eta_{33}$ are consistent with equation \ref{eq:lambda33lin}.

From these expressions, we can check the consistency of the approach by predicting the steady state volume fraction profiles in the two resuspension geometries. For that purpose, we simply integrate equation \ref{eq:sbmdimensionless}, using the above expressions and imposing the volume of particles to set the integration constant. The resulting volume fraction profiles are superimposed to the numerical data in Fig. \ref{fig:phi_profiles}. The agreement is rather good, except for the highest Shields numbers. This is due to the fact that the volume fraction profiles in viscous resuspension are more sensitive to the particular form of the empiric functions chosen for $\eta_{ii}$. Indeed, the normal viscosity coefficients are plotted in Fig. \ref{fig:eta_model} in log scale over 4 orders of magnitude.

\subsection{Comparison with experiment: role of inertia}
\label{subsec:inertia}

Resuspension in the shear direction with a uniform shear rate is not documented in the literature until now, and it is probably difficult to perform it experimentally. Direct comparisons with three available sets of experiments \citep{Acrivos1993,SaintMichel,Dambrosio} are possible for resuspension in the vorticity direction, and are presented in this section. We first focus on the height increment induced by resuspension. 

\begin{figure}
    \centering
    \includegraphics[width=0.7\linewidth]{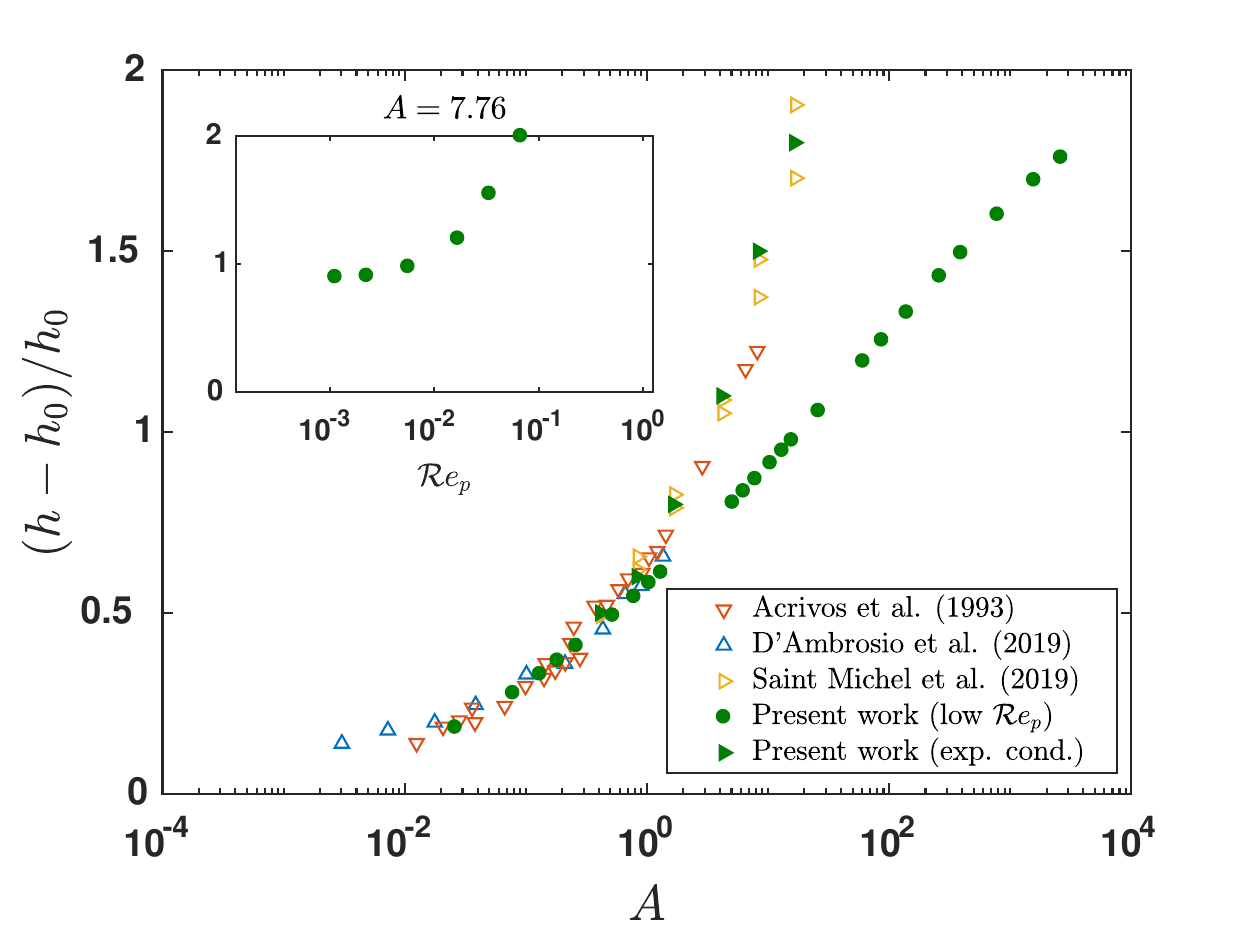}
    \caption{Comparison of measured height from Couette resuspension experiments (open symbols) and from simulations (filled symbols). The triangular filled symbols are simulations using the same parameters as in \citep{SaintMichel} experiments. Insert shows simulations results done at fixed Acrivos parameter, varying the particles Reynolds number.}
    \label{fig:h_h0}
\end{figure}

In Figure \ref{fig:h_h0} the height increment is presented as a function of the Acrivos parameter $A$, defined by $A=9/2(\eta_f\dot\gamma)/(\Delta\rho g h_0)$, for the sake of consistency with earlier authors. $A$ is proportional to the Shields number ($A= 9/2(a\Sh)/h_0$). The simulation results are in excellent agreement with the available data For $A<2$ ($\Sh<20$). In contrast, a significant discrepancy is found for higher values of $A$, when the mean volume fraction of the bed is less than 35\% approximately. 

There can be two main reasons for this discrepancy. The first one is coming from the interaction model used in the simulations. Indeed, the viscosity of the suspending fluid is accounted for in the particle interactions via the lubrication terms exclusively while other, non-diverging, hydrodynamic interactions certainly matter in sufficiently dilute regimes. However, previous results \citep{chevremont2019} concluded that, in terms of bulk viscosity, the agreement between well resolved Stokes solutions and a simple lubrication model was rather good down to approximately $\phi=0.2$. It is therefore unlikely that the limits of the models are the main reason for discrepancies which appear at $\phi=0.35$. 

The second possible reason concerns the experimental parameters used in reference \citep{Acrivos1993} and \citep{SaintMichel}. In the simulations, we paid particular attention to avoid any inertial effects. In  particular, the particles Reynolds number, 
$\mathcal Re_p = \rho \dot\gamma a^2/\eta_f$ is always below $5~10^{-3}$
In the experiments, it is not always the case. In particular, we found that the few experimental points which deviate from the simulation results  ($A>2$) correspond to particles Reynolds numbers of the order of, or greater than 0.01. We performed additional simulations in this regime (inserts of figure \ref{fig:h_h0}), and observed that the resuspension was significantly enhanced. Moreover, when using similar parameters as in \citep{SaintMichel}, the simulation data capture very well the reported suspension height. We can thus conclude that the reason for the apparent discrepancy at high Shields number is due, mainly, to the fact the inertia was not negligible in the experiments. When both simulation and experiments are in the Stokes limit they are in very good agreement.  

The onset of inertia enhanced resuspension is about $\mathcal Re_p \sim  10^{-2}$, though the effect of inertia becomes non-negligeable from $\mathcal Re_p \sim 5\times 10^{-3}$. This number might appear quite small. Transition from viscous to inertial regimes, though described by Bagnold decades ago \cite{bagnold1954experiments} is still in debate in the recent literature \cite{madraki2020shear,tapia2022viscous}. A few numerical work \cite{trulsson2012transition,amarsid2017} report a transition value around unity, whereas experimental work report much lower values, especially at high volume fractions \cite{fall2010shear, madraki2020shear}, with the exception of the recent work of Tapia and co-workers which reports a transition $\mathcal Re_p$ above unity. The above mentioned discrepancies might result from the role of friction \cite{tapia2022viscous,degiuli2015unified}: frictionless systems may exhibit a transition at lower shear rates. Let us recall that here, contacts are frictional. It should also be noticed that - except the present work - simulations have been carried out in 2D. It has been recently argued \cite{madraki2020shear} that the relevant Reynolds number should be based on the local shear rate (i.e. the one one in between particle pairs), which is greater than the mean shear rate by a factor $F(\phi)$ which is an increasing function of the volume fraction. In the work of Madraki and co-workers \cite{madraki2020shear}, transition from viscous to inertia was found to be as low as $\mathcal F(\phi) \mathcal Re_p \sim \times 10^{-1}$. Here, when the relative height increment is of the order of unity, the mean volume fraction of the bed is about 30\%, and we estimated $F(\phi) \mathcal Re_p \sim 8\times 10^{-2}$ (using expression given in reference \onlinecite{madraki2020shear}), a value which is rather close to the one found experimentally by Madraki and co-workers. The present low $\mathcal Re_p $ found at the transition is therefore in agreement with the experiments reported in reference \onlinecite{madraki2020shear}, but not with those in reference \onlinecite{tapia2022viscous}. Though a dedicated study seems necessary to shed some light on this debate, we can however conclude that inertia effects appear at rather low Reynolds number in viscous resuspension. Part of the existing experimental results on viscous resuspension were in fact crossing the viscous to inertial transition (in ref. \onlinecite{Acrivos1993} and \onlinecite{SaintMichel}) when increasing the Acrivos number, and should therefore be handled with caution. 

In the experimental works reported in \citep{SaintMichel} and \citep{Dambrosio}, not only the height increment but also the profiles of volume fraction were determined. The normal viscosity coefficient $\eta_{33}$ deduced in \citep{SaintMichel} from these profiles matches equation \ref{eq:normvisc_correlations} with \citep{Boyer2011b}'s parameters. Conversely, \citep{Dambrosio} suggest a better agreement with \citep{Zarraga2000}'s one. The present results highlight that the superior accuracy of one expression entirely depends on the range of $\phi$ considered. Overall, Boyer's expression is suitable for $\phi/\phi_m>0.8$ and \citep{Zarraga2000}'s one for $\phi/\phi_m \in [0.6, 0.8]$. They both overestimate the viscosity if $\phi/\phi_m < 0.6$ . These discrepancies deserve several comments. In both experimental works, the normal stress was found to be shear-thinning, which is not found with the present model. It is thus clear that the experimental systems are physically more complex. The fact that the low values of the normal stress that we report for $\phi<35\%$ were not observed in experiments may come in part from inertial effects, since low volume fractions are mainly obtained at high Acrivos numbers where the Reynolds number was not kept small in the experiments. 

Another clear difficulty encountered in this work is related to the interpretation of the results in the regions with strong gradients of solid fraction. Gradient effects can be expected in experiments too, and the only way to reproduce them numerically would be to replicate the particles in terms of both size and number - which we did not do as it would have required extremely large numbers and prohibitive computation times. Nevertheless, the numerical results give an insight into gradient effects: when the gradient of $\phi$ is greater than a fraction of the inverse of the particle size, it is not possible to collapse the profiles obtained at different Shields numbers.

\section{\label{sec:conclusion}Conclusions}

We reported in this paper numerical data on viscous resuspension of buoyant particles in both the vorticity and the shear direction, together with direct determination of normal stresses in homogeneous shear flow of non-buoyant particles. Resuspension profiles are consistent with the normal stresses obtained in homogeneous conditions. At steady state, weight is balanced by the divergence of a stress to which both contacts and lubrication contributes. Resuspension simulations allow to obtain very efficiently a large and robust set of data but we emphasized that on the top of the resuspended bed, strong volume fraction gradients cannot be accounted by a continuum approach.  


The accuracy of the results and the large range of volume fraction investigated let empirical expressions be proposed for the normal stress components. These expressions differ significantly from the ones proposed earlier, both qualitatively and quantitatively. First, the normal stress ratio are not constant, they greatly depend on the volume fraction. In the semi-dilute regime, the normal stress is much larger in the shear direction than in the vorticity direction. Conversely, a spherical tensor is enough to describe the normal stresses in the concentrated regimes ($\sigma_{11} \simeq \sigma_{22} \simeq \sigma_{33}$). Second, none of the earlier proposed expressions can account for the normal stress components in the full range of volume fraction. In particular we find that  $\sigma_{22}$ and $\sigma_{33}$ decrease sharply below $\phi\sim 0.35$, when contacts and lubrication contributions tend to cancel out. These features were evidenced by the numerical simulations and are reflected in the new empirical relations.

Though mainly focused on the viscous limit, we also observed inertial effects when the particle Reynolds number was overset above $5\times 10^{-3}$. The resuspensed height of the bed increased significantly as compared to the viscous limit, i.e. effective normal stress was increased due to inertia. Part of the available experimental data in the literature were obtained above this limit at high Acrivos numbers. Therefore, previously reported relations for $\sigma_{33}$ need therefore to be taken with caution, especially at low volume fractions. Above $\mathcal{R}e_p \sim 5\times 10^{-3}$, normal stress are functions of both $\phi$ and $\mathcal{R}e_p$ and both were varied at the same time in the experiments. This observations partly explained the discrepancy between experimental results, and clearly ask for systematic studies in order to extend normal stress models to moderate Reynolds numbers.    


We only tested homogeneous shear rates configurations, and it seems important to extend this approach to non-homogeneous ones as in many situations (e.g. Poiseuille flow, sediment load), there exists gradient of shear rate which can have consequences on particle migration. Besides, as many recent experimental results report a non-linear dependency of the stress with respect to the shear rate, it seems to be important to incorporate in numerical models additional phenomena such as a variable friction coefficients, long-range interactions or non-Newtonian fluid model.

\section*{\label{sec}Acknowledgements}
Most of the computations presented in this paper were performed using the GRICAD infrastructure (https://gricad.univ-grenoble-alpes.fr), which is supported by Grenoble research communities.

\bibliographystyle{unsrtnat}
\bibliography{biblio}
\end{document}